\documentclass[manuscript]{acmart}

\AtBeginDocument{%
  }

\setcopyright{none}
\usepackage{tabularx}
\begin{document}

\title[From AI-Free Gates to AI-First Work]{From AI-Free Gates to AI-First Work: How Engineering Students in India Navigate University Education and Career Preparation Without AI Guidelines}

\author{Samhitha Goutham}
\affiliation{%
  \institution{Independent Researcher}
  \city{Chennai}
  \country{India}}
\email{samhithagoutham24@gmail.com}

\author{Ziwei Gu}
\affiliation{%
  \institution{Harvard University}
  \city{Cambridge}
  \state{Massachusetts}
  \country{USA}}
\email{ziweigu@g.harvard.edu}

\renewcommand{\shortauthors}{Goutham and Gu}

\begin{abstract}
Engineering graduates in India enter a technology sector that is being reshaped by generative AI, yet the route into that sector runs through examinations and campus placement screens that largely exclude AI. Institutions in India have an approved intake of nearly 2.9 million seats in engineering and technology for 2026–2027~\cite{aicte2026intake}, but HCI research on how students prepare for AI-mediated work has drawn mostly on Western and East Asian settings. We interviewed 23 students and recent graduates of Indian engineering and computing programmes about how they used generative AI across coursework, examinations and placements, and how they judged their own readiness for work. Using reflexive thematic analysis, we report five themes, covering AI-first workplaces and AI-free placement screens, how participants prepared for placement, how they used AI in coursework, the role of faculty authority and enforcement, and participants’ concerns about their readiness for work. Drawing on signaling theory, we discuss what grades and screen scores certify once AI can help produce them, and we offer recommendations for employers, institutions and AI tool designers.
\end{abstract}



\maketitle

\section{Introduction}\label{sec:intro}

Many students in Indian engineering programmes hope to join the technology sector, whether IT services firms, which give India a large share of the world’s outsourced technology services~\cite{nasscom2025strategic}, or product companies and large technology firms in India and abroad. The sector is itself being reshaped by generative AI (GenAI). In software engineering more broadly, work that once trained junior engineers is being redirected into workflows that pair senior engineers with AI~\cite{yu2026seniorgenerativeaierodes}. The route into this work, however, runs through a series of high-stakes gates, most of which still assess students without AI. Admission rests, depending on the institution, on national, state or institution-specific entrance examinations or on marks in the 12th-grade board examinations. Once enrolled, students are graded through semester-wise examinations and other forms of assessment, which together contribute to a cumulative grade point average (CGPA). In the later years, campus placements, in which companies recruit directly from institutions, screen students through CGPA cut-offs, online aptitude and coding tests and interviews (Table~\ref{tab:terms}). Teaching is often lecture-centred, and assessment in this legacy system relies heavily on theory and rote learning.

Students in this system speak of \emph{clearing} an examination, the word Indian students and universities commonly use for passing it. Our participants used the word in a sharper sense, for getting through an assessment with the marks it requires, and they set it against learning what the assessment was meant to teach. GenAI tools such as ChatGPT, Gemini and GitHub Copilot entered this environment without institutional frameworks for their use. Students were left to decide for themselves how to use the tools to clear coursework, how to prepare with them for placement screens that mostly forbade them, and how to become ready for workplaces that would expect them.

Research in HCI and CSCW has begun to trace how GenAI is changing the transition from education into technical work, from the junior roles it absorbs~\cite{yu2026seniorgenerativeaierodes} to the automated screens that filter applicants~\cite{10.1145/3711038}. Most of this work, like research on student AI use more broadly, draws on participants in North America, Europe, Australia and East Asia~\cite{10.1145/3706598.3713393,10.1145/3724363.3729024,10.1080/02602938.2025.2502582,yu2026seniorgenerativeaierodes}. The closest studies of Indian engineering students date from shortly after ChatGPT’s release and focus on how students used the tools and how the tools could be improved~\cite{10.1145/3636243.3636257,joshi2023withgreatpowercomes}. Neither examined campus placements, although placements organise much of the later years of an Indian engineering degree and decide where many graduates begin their working lives. Less is known about how students move through these gates now that AI use has become routine, and about what they carry from them into AI-first work.

We draw on signaling theory~\cite{10.2307/1882010} to interpret this transition. Grades and screen scores are signals that employers rely on because they cannot observe a candidate’s ability directly (Section~\ref{sec:signal}), and we ask what these signals certify once AI can help students produce them.

To understand this transition, we ask three research questions:

\begin{itemize}
  \item \textbf{RQ1.} How do engineering and computing students and recent graduates in India perceive their readiness for AI-first work, and how do they prepare for it?
  \item \textbf{RQ2.} How do they use generative AI to clear the coursework, examinations and placement screens that lead to that work?
  \item \textbf{RQ3.} How, in their accounts, do the institutions around them shape that use?
\end{itemize}

We address these questions through semi-structured interviews with 23 current students and recent graduates of engineering and computing programmes. Many participants began college around the time ChatGPT was released and have graduated, or will soon graduate, into workplaces that expect AI use, while others entered college with AI already part of their schooling.

We report five themes. 1) Placement screens excluded AI, while workplaces expected it. 2) Participants prepared for placement with AI as their coach. 3) They used AI to clear coursework, and sometimes to learn. 4) Faculty set the rules, while peers taught one another how to use AI. 5) Participants doubted their readiness for AI-first work.

This paper makes three contributions to research on work and its futures. First, it offers an empirical account, based on interviews conducted in 2026, of how students in an underrepresented setting move from education into AI-first technical work, centred on the gates between the two. Second, it uses signaling theory to show how AI-assisted clearing may weaken the grades and screen scores through which employers select graduates. Third, it identifies the responsibilities students take on when neither institutions nor employers define what readiness for AI-first work requires, and it derives implications for employers, institutions and AI tool designers.

\begin{table}[tp]
  \caption{Indian engineering education: terms used in this paper.}
  \label{tab:terms}
  \small
  \begin{tabularx}{\textwidth}{@{}>{\raggedright\arraybackslash}p{3.6cm}X@{}}
    \toprule
    \textbf{Term} & \textbf{Meaning} \\
    \midrule
    Institution types & Engineering and computing degrees are offered by centrally funded technical institutes, state public universities, private universities, private colleges affiliated to a university, and arts-and-science colleges offering B.Sc. computing degrees. Admission routes vary. Some institutions admit through national or state entrance examinations, some through their own entrance tests and others through 12th-grade board examination marks, and institutions are informally ranked by how selective their admission is. \\[3pt]
    Board examinations & Examinations at the end of the 12th grade, the final year of school, set by state or national school boards. Some institutions admit students on their marks in these examinations. \\[3pt]
    Semester examinations & Examinations at the end of each semester, mostly closed-book and handwritten, set and marked within the institution. They usually carry a large part of a course’s grade. \\[3pt]
    Two-mark and five-mark answers & Question papers state the marks for each question, which tells students how long and how detailed an answer is expected to be. \\[3pt]
    Internals & Internal assessment marks from assignments, class tests and lab work, counted toward the course grade alongside the semester examination. \\[3pt]
    Lab record & A notebook of laboratory exercises submitted for marks. \\[3pt]
    Viva & A viva voce, or oral examination, in which students explain their work to an examiner. \\[3pt]
    Mug up & Colloquial for memorising material in order to reproduce it in an examination, usually without understanding it. \\[3pt]
    CGPA & Cumulative grade point average across all semesters, typically on a 10-point scale. \\[3pt]
    Campus placements & Recruitment drives that institutions organise in students’ pre-final and final years. Each company sets its own screens, usually a minimum CGPA, online aptitude and coding tests, sometimes group discussions, and technical and HR interviews. \\[3pt]
    DSA & Data structures and algorithms, the most demanding part of most coding tests and technical interviews. Screens also cover quantitative and logical aptitude, core subjects of the degree and other topics, but participants spoke mostly of DSA. \\[3pt]
    \bottomrule
  \end{tabularx}
\end{table}

\section{Related Work}\label{sec:related}

Our study draws on three lines of research. We first review work on how GenAI is changing the transition from education into technical work, including hiring, and introduce the signaling lens we use. We then turn to studies of GenAI use by students in India and the Global South, and to research on how assessment, enforcement and reliance shape students’ use of AI. We close by identifying the gap this study addresses.

\subsection{From Education to AI-First Work}\label{sec:signal}

Recent work traces how GenAI is reshaping the path into technical work. From interviews with eight Korean engineering and computer science students and recent graduates and six senior software engineers, Yu and Moon describe a process they call \emph{Absorption}, in which GenAI redirects entry-level work into workflows shared by senior engineers and AI, and removes the productive struggle through which juniors once built expertise~\cite{yu2026seniorgenerativeaierodes}. Their juniors reported grades that stayed stable while learning diverged, and a curved grading system made opting out of AI costly. Research on knowledge work points to a related risk. Surveying 319 knowledge workers, Lee et al. found that those with more confidence in GenAI reported engaging in less critical thinking, particularly when tasks seemed trivial, when time was short or when they lacked the knowledge to evaluate the output, and they warn that skills left unpractised on routine tasks may deteriorate~\cite{10.1145/3706598.3713778}.

Hiring is changing too. Surveying 448 computer science students at three Philadelphia universities, Armstrong and Metaxa found that students saw automated screening as less fair than human review, and that only referrals and family household income predicted job offers, whereas more egalitarian strategies such as free coding-assessment practice did not~\cite{10.1145/3711038}. In India, the route into technical work runs largely through campus placements, which institutions organise but in which each recruiting company screens candidates in its own way (Table~\ref{tab:terms}).

Signaling theory offers a way to read such gates. Spence modelled hiring as a decision under uncertainty, in which employers cannot observe a candidate’s productivity and instead infer it from observable signals such as education, which candidates choose to acquire at a cost~\cite{10.2307/1882010}. A signal separates candidates only if it is costlier to acquire for those who are less productive, and employers’ beliefs about what it indicates are sustained only as long as hiring outcomes confirm them. CGPA cut-offs and standardised placement tests operate as signals of this kind, applied at scale to cohorts that recruiters cannot assess individually. We ask what happens to these signals when AI can help students produce them.

\subsection{Students and GenAI in India and the Global South}

The closest prior work studied the same population shortly after ChatGPT’s release. Budhiraja et al. surveyed 480 computer science undergraduates at three Indian institutes and interviewed 17 students from eight universities~\cite{10.1145/3636243.3636257}. Students used ChatGPT for information gathering, summarising and coding, but most worried about incorrect output and their inability to verify it. In a preprint from the same group, Joshi et al. extended this work to 1,306 survey responses and interviews with 112 students and 27 instructors at three Indian universities~\cite{joshi2023withgreatpowercomes}. Some of their instructors had already withdrawn open-book assessments and expected closed-book examinations to remain important, and the instructors were divided on whether using ChatGPT to prepare teaching material was legitimate. Both studies date from 2023 and centre on usage, perceived benefits and challenges, and how the tools could be improved.

Beyond India, research on AI in Global South education has focused largely on school-age learners. Verma et al.’s review of 18 studies of child–computer interaction with AI-based educational tools in the Global South, in which India accounted for the largest share, identifies limited access to technology and models trained mainly in English among the main challenges~\cite{10.1145/3736645}. In an experiment with 118 participants from India and the United States, Agarwal et al. found that AI writing suggestions gave Indian participants smaller efficiency gains and moved their writing toward Western styles~\cite{10.1145/3706598.3713564}. Sebastian et al.’s design workshops with students in Finland and India attribute differences in how students oriented to AI to the pedagogical ecologies of their schools more than to national cultures~\cite{10.1145/3772318.3793808}. We adopt the same emphasis and attend to institutions instead of an essentialised “Indian culture.” Following Cao et al., who argue for building AI literacy education on what learners already bring instead of on what they lack~\cite{10.1145/3706598.3713173}, we also attend to the strategies participants developed without institutional guidance.

\subsection{Assessment, Enforcement and Reliance}

Assessment research offers a lens on how institutions respond to GenAI. Corbin et al. distinguish \emph{discursive} changes to assessment, such as rules and declarations about permitted AI use, from \emph{structural} changes to the task itself, such as supervised work and interactive oral assessment~\cite{10.1080/02602938.2025.2503964}. Because AI detection is unreliable, they argue, frameworks that rely on discursive rules create only an illusion of enforcement. In a small Swedish study of a course that paired an AI-permitted take-home examination with a viva, Lindkvist et al. found that, for the two students who chose not to use GenAI, the viva worked as a check that kept their own understanding in view~\cite{10.3389/feduc.2026.1813306}.

Where rules exist, students navigate them strategically. Adnin et al. document how U.S. students conceal GenAI use by paraphrasing output or adjusting tone, and calibrate this effort to each instructor’s perceived strictness~\cite{10.1145/3706598.3713393}. Concealment also complicates measurement. Ling et al. found that about 60\% of the undergraduates they surveyed reported using AI themselves while nearly 90\% said their peers did, a gap students attributed to embarrassment~\cite{10.1145/3772318.3791073}, which informed our decision to ask participants about their peers as well as themselves. Where AI use is permitted, students decide where in a task to involve it, and Bodei et al. found that most students in a UK programme that permitted and taught AI use involved it in only one or two of five writing stages~\cite{10.1145/3808045.3808072}.

GenAI is also changing whom students ask for help and how much they rely on it. Hou et al. found that help-seeking among North American computing students had become AI-mediated, with some reporting isolation and shame about using AI in front of peers~\cite{10.1145/3724363.3729024}, and Henderson et al. found that Australian students trusted teacher feedback far more than GenAI feedback but saw GenAI as less risky to approach~\cite{10.1080/02602938.2025.2502582}. Reliance research suggests that novices are poorly placed to notice when AI is wrong. Bo et al. found that most undergraduates in their experiment did not notice a chatbot’s sycophancy~\cite{10.1145/3772318.3791365}, and Ma et al. found that prior experience remained a significant predictor of success in a data-science course when students had ample time, even with LLM assistance~\cite{10.1145/3772318.3791283}. Ammari et al. propose \emph{repair literacy}, the competencies students develop to diagnose and recover from AI failures, as a form of literacy learned through use~\cite{ammari2026repair}.

Research on how GenAI is changing the move into technical work comes mostly from the United States and Korea, and research on assessment, disclosure and help-seeking mostly from settings where AI use was permitted or taught~\cite{10.1145/3808045.3808072,10.3389/feduc.2026.1813306}. Existing studies of Indian engineering students date from 2023, focus on the tools and do not examine placements~\cite{10.1145/3636243.3636257,joshi2023withgreatpowercomes}. No prior study follows students through an exam-oriented system with steep faculty hierarchies and little stated policy, past placement screens that largely exclude AI, and into workplaces that expect it. We address this gap through interviews with students and recent graduates whose education spanned the arrival of GenAI.

\section{Methods}\label{sec:methods}

\subsection{Participants and Recruitment}\label{sec:participants}

This study uses a qualitative design, with semi-structured interviews as the primary method of data collection, because our questions concern how participants experienced and reasoned about AI use, which requires interpretive depth more than numerical breadth. Participants were recruited through purposive sampling, supplemented by snowball sampling through participant referrals. Participants were required to be 1) current students, or 2026 graduates, of undergraduate or integrated engineering or computing programmes at Indian colleges or universities, 2) current or past users of AI tools in their academic work, and 3) able to take part in an interview in English or Tamil. Including recent graduates allowed us to hear from participants whose degrees spanned the release of ChatGPT and who had since entered work. Recruitment through referrals may explain why the sample clusters in a few institutions, and we avoid institution-level claims that rest on one or two accounts.

Of the 23 participants, 13 were current students and 10 had graduated. Two participants were women. Thirteen studied in B.E. or B.Tech. programmes across eight engineering disciplines, and the remainder in integrated M.Sc. or B.Sc. computing programmes. Their institutions included two centrally funded technical institutes, several private universities, a public state university, a private engineering college and an arts-and-science college. Most participants studied in \anon[one South Indian city]{Chennai}, and all were from established institutions of good standing. Appendix~\ref{app:participants} lists participants’ attributes. To protect anonymity, we refer to participants only by ID (P1–P23) and use the singular “they” for all of them.

\subsection{Data Collection}

The first author conducted all interviews, in person or by video call, between September and October 2026. Seventeen interviews were conducted in English. In the other six, participants preferred to speak partly or mainly in Tamil, and the interviewer conducted those exchanges in Tamil. Most interviews lasted between 30 and 45 minutes, and a few Tamil–English interviews were shorter, at about 20 minutes. The guide comprised 21 questions in seven sections, covering 1) background and academic context, 2) AI use in coursework and study, 3) AI use for examinations and placements, 4) perceptions of learning and skill development, 5) disclosure and institutional norms, 6) technical readiness and career concerns, and 7) reflections and recommendations. It was reviewed with the second author before data collection, and questions were occasionally omitted when a participant had already addressed them. When a participant struggled to phrase an answer, the interviewer sometimes summarised what they had said or offered possible wording. We did not use quotes that only agreed with an interviewer’s suggestion, and we indicate in the Findings where a reported view followed such a prompt.

\subsection{Transcription, Translation and Analysis}

Interviews were transcribed using automatic speech recognition (noScribe). Where a quoted word had clearly been misrecognised, particularly product names and technical terms, we corrected it after checking the context and the recording, and we otherwise retained participants’ wording, including disfluencies that do not obscure meaning. In quotes, square brackets mark words added or replaced for clarity, and […] marks omitted text. Tamil–English segments were translated into English by the first author, a native Tamil speaker. Translated quotes are marked with †.

We followed the six phases of reflexive thematic analysis~\cite{braun2006using}. The first author coded the transcripts, developed the themes and kept reflexive memos. Coding was inductive and proceeded in two levels. The first author first coded every transcript and developed first-level themes, in two batches as interviews were completed (the first 15, then the remaining eight, including all six Tamil–English interviews). These first-level themes were then grouped into broader second-level themes, which formed around a recurring tension between the understanding participants valued and a system that, in their accounts, rewarded \emph{clearing}. For this paper, we organised the themes around participants’ movement from coursework into work, consolidating related themes into five. We treat themes as interpretive constructs, not frequency counts, and list by ID the participants who described each pattern so that readers can judge how widely it was shared.

\subsection{Positionality}

The first author graduated in computer science in 2024 from one of the institutions represented in the sample, works as an AI consultant and conducted all interviews. This familiarity with Indian engineering education and industry practice informed the interview guide and helped participants speak openly, but it also carries the risk of assumptions shaped by the first author’s own trajectory. To limit this, the first author kept reflexive memos and consulted regularly with the second author, a PhD student in human–computer interaction who did not take part in data collection, and we grounded each claim in participants’ own words.

\subsection{Ethics}

The study was approved by an institutional review board (IRB). All participants were 18 or older. Before each interview, participants were told the study’s purpose, their right to withdraw at any time without penalty and how their data would be handled, and they gave informed consent, including consent to audio recording. Recordings and transcripts are accessible only to the two authors. Participation was voluntary. Institutions are described by type, and professors’ names mentioned by participants were removed from transcripts.

\section{Findings}\label{sec:findings}

We present five themes that follow participants along the path from coursework to work (Table~\ref{tab:themes}). We begin where that path ends, with the AI-first workplaces graduates described and the placement screens that stand before them (Section~\ref{sec:t1}), and with how participants prepared for those screens with AI as their main coach (Section~\ref{sec:t2}). We then turn back to the coursework whose grades the screens filter on, which participants used AI to clear but which also shows how AI helped them learn (Section~\ref{sec:t3}), to the faculty authority and enforcement that shaped how openly they did so (Section~\ref{sec:t4}), and finally to participants’ concerns about what all of this left them ready for (Section~\ref{sec:t5}).

\begin{table}[tp]
  \caption{Themes, a representative quote for each, and the research questions they address.}
  \label{tab:themes}
  \small
  \begin{tabularx}{\textwidth}{@{}>{\raggedright\arraybackslash}p{4.4cm}Xl@{}}
    \toprule
    \textbf{Theme} & \textbf{Representative quote} & \textbf{RQ} \\
    \midrule
    4.1~AI-Free Gates, AI-First Work & “\emph{we’re meant to do it AI first because, uh, we’re always meant to deliver results as quick as possible.}” (P9) & RQ1, RQ2 \\[4pt]
    4.2~Preparing for the Gate with AI & Professors “\emph{come to teach their subjects, they will move on.}” (P14) & RQ1, RQ3 \\[4pt]
    4.3~Clearing, Not Learning & “\emph{So, we clear the exams but not the learning part. We never get it.}” (P13) & RQ2 \\[4pt]
    4.4~Answering Upward, Sharing Sideways & “\emph{I believe the professor more. Because, again, the professor is the one giving me the marks, right?}” (P8) & RQ2, RQ3 \\[4pt]
    4.5~Ready for What? & “\emph{you don’t go with the same ownership confidence that you used to do.}” (P7) & RQ1 \\[4pt]
    \bottomrule
  \end{tabularx}
\end{table}

\subsection{AI-Free Gates, AI-First Work}\label{sec:t1}

For most participants, the assessment that mattered most was campus placement (Table~\ref{tab:terms}), and it was where their use of AI was most intensive and most purposeful. Yet in their view the placement process tested skills they did not expect to use, in a form that excluded the AI their future workplaces would require. Graduates’ accounts of those workplaces, and of how the standing of AI shifted while they were still in college, sharpen this contrast.

\paragraph{Workplaces that expect AI first}
Graduates described workplaces in which AI was the default. At the company where P9 now worked as a junior developer, “\emph{we’re meant to do it AI first because, uh, we’re always meant to deliver results as quick as possible,}” and project deadlines that would once have run to a month had shrunk to one or two weeks. P17, still a student, had heard that employees in companies were “\emph{forced to use only AI. They are forced to use [vibe] coding,}” and worried that this would erode their coding skills. P5, a biomedical engineering graduate now in work, said that “\emph{everything is just dumped on AI.}” Internships carried the same message back to campus. P8’s managers at an internship confirmed what P8 had come to believe as models improved, that “\emph{if you don’t use AI, you’re probably falling behind,}” and for P10 the question was settled. “\emph{The world’s moving forward. We should move forward.}”

\paragraph{From cheating to expectation within one degree}
Most graduates had entered college around 2022, as ChatGPT was released, so the norms around AI changed while they were still studying (P2, P4, P6, P8–P10, P12, P14, P15). In P9’s first year, language models were still “\emph{gimmicky toys that you would try and have fun with,}” and using AI to write code “\emph{was viewed as a kind of a cheaty way to go about things […] a way that you’ve like beaten the system.}” By the end of their degrees, as P4 put it, “\emph{by the third or fourth year, AI basically took over everything.}” Faculty attitudes moved in the same direction (P8, P10, P14). P8 recalled professors who had been “\emph{quite serious}” about banning AI in the first two years but who, “\emph{as the years went on,}” “\emph{just, like, looked the other way when they noticed what there’s AI in it,}” and P14 recalled professors who had been “\emph{strictly against AI}” later using it themselves. P9 judged the absence of guidelines during this period understandable, since “\emph{the entire situation around LLMs was pretty much unknown and guidelines wouldn’t have served the purpose right at that time.}”

Participants’ own unease often faded with these norms. P9 had once felt that using AI on an assignment was “\emph{pointless}” when it bypassed what the assignment was meant to teach, but now saw learning as “\emph{almost, uh, like secondary now because of how like AI native everything has become.}” The shift was not universal, and P2 and P7, both graduates, still spoke of guilt and of not fully owning AI-assisted work (Section~\ref{sec:t5}).

\paragraph{A gate that tests something else}
Against this picture of work, many participants felt that placement screens tested something other than the job (P3, P4, P9, P15, P16). P3 argued that aptitude tests screened out capable people:

\begin{quote}
\emph{Some person could be, like, very strong in the actual job rather than, like, being smart in, like, an aptitude test. So, he fails in that and somebody who is, like, not smart enough for the job but is smart enough for the aptitude could clear the aptitude and get the job.} (P3)
\end{quote}

P15 estimated that about half of the companies tested only aptitude, looking for “\emph{that sharpness or what, presence of mind,}” while P16 had faced DSA questions for a front-end role and P4 found that mechanical engineering interviews asked theoretical questions unrelated to field work. To P9, interviews had become “\emph{a lot harder than what the actual job might be,}” since DSA concepts “\emph{don’t really come up when you’re actually tasked to do your job.}”

Participants rarely objected to the gate itself, however. Several accepted it as a filter for general ability, not a test of job skills (P5, P9, P10). Given “\emph{the sheer volume of candidates,}” P9 reasoned, the screens identified those “\emph{with the problem solving mindset that who can later be trained,}” and P10 agreed that companies “\emph{want to check your raw, you know, problem solving skills}” because “\emph{they’re going to take you and they’re going to teach you anyways.}” P6 was alone in finding that interview questions matched their daily work.

\paragraph{Screens that exclude AI, and do not hold}
Most placement screens assessed candidates without AI, and participants doubted that this held in practice. P8 believed that “\emph{80 to 90\% people cheat}” in online placement tests, which in P8’s view left selection close to chance. P5 was an exception who had made a deliberate choice, refusing on principle to use AI on online assessments because P5 held that their results should reflect what they actually knew. A few companies had adapted, and participants welcomed them. P1 recalled online assessments that allowed Copilot, which “\emph{would give us the syntax and stuff, but it wouldn’t give us the logic,}” and argued that this “\emph{should be normalized.}” P8 described companies that replaced online tests with building an application, with AI permitted, as “\emph{more of a reflection of what’s going on in real life.}”

Participants thus prepared for screens that excluded AI in order to enter workplaces that required it. How they prepared, and with whose help, is the subject of the next theme.

\subsection{Preparing for the Gate with AI}\label{sec:t2}

Participants treated placement as the real target of their learning, and their accounts of preparing for it centred on their own study, their peers and AI. Faculty appeared in these accounts mainly as teaching subjects that the screens did not test, and only occasionally as a source of placement advice (P13). Almost every participant used AI to prepare for placements or job interviews, all except P11, a first-year student, and they did so with a sophistication rarely visible in their accounts of coursework.

\paragraph{Placement as the target of learning}
In participants’ accounts, the curriculum did not prepare them for what placement screens tested. Professors did not cover the DSA problems that dominate coding tests, P14 explained, “\emph{because they come to teach their subjects, they will move on,}” so P14 asked AI to plan a preparation schedule and to test them along the way. P18, a final-year computer science student, drew the line between coursework and placement before any summary from the interviewer:

\begin{quote}
\emph{But really I didn’t learn a lot from that. The main intention for me was to get marks and CGPA. But then as for now it’s really important to get into a job for for that I should learn this coding data structure.} (P18)
\end{quote}

Placement also reached back into coursework. Because companies set “\emph{a minimum requirement of CGPA, which is, like, mostly high,}” P1 explained, students used AI to keep their grades up at the expense of time for projects. The CGPA that screens filtered on was thus partly produced with the same tools.

\paragraph{AI as placement coach}
Participants used AI to rehearse the process itself, not only to learn its content. Several ran mock interviews with AI (P3, P6, P10, P12, P22, P23). P6 told the model to order questions by difficulty and to “\emph{be brutally honest, and you know, kind of grade my answer,}” using it to check whether they were “\emph{on point}” and “\emph{not waffling too much,}” while P12 rehearsed answers by voice and had AI flag parts that were “\emph{unnecessary}” or “\emph{too personal.}”

Preparation was tailored to the company, the role and even the interviewer. Before one interview, P5 pasted the job link and the LinkedIn profiles of both interviewers into ChatGPT and asked, “\emph{what are their specialties? What are they most likely to ask questions about?}” P9 similarly anticipated questions “\emph{from the perspective of this interviewer who is interviewing for this particular job description.}” Some participants built tools of their own. P21 created a custom GPT to act as “\emph{my aptitude trainer, and me as its student,}”† P17 asked ChatGPT for the problems particular companies, such as Amazon and Flipkart, asked most often, “\emph{instead of, uh, blindly, uh, solving problems using leet code,}” and P19 fed the URLs of aptitude question sites to AI to generate timed practice tests. AI also built participants’ applications. P21 said that the resumes submitted to three or four companies were “\emph{all […] resumes built with AI,}”† while P13 tailored a resume to each role on a faculty member’s advice, making sure that it stayed “\emph{at least a bit true, not a false information.}”

Much of this preparation was compressed into very little time. Shortlists were often released the day before an interview, P8 explained, “\emph{so, like, I’ll have to prepare overnight,}” and AI provided “\emph{a good consolidation of what I should be learning.}” P14, with no time to spend two or three days on tutorials, asked AI to teach the JavaScript concepts they could learn “\emph{within two hours to attend the interview.}” P6 doubted they could have managed without it. Preparing for an interview without AI, they said, “\emph{I think I’d have a panic attack straight up.}”

\paragraph{Unequal access}
The quality of this coaching depended partly on money. No participant said their institution provided AI tools directly (P1, P3–P5, P10–P12, P14, P16–P19, P21–P23), and access came instead from companies’ student promotions and telecom bundles (P1, P2, P4, P10, P12, P14–P20). Free tiers sufficed for coursework, P7 observed, because the syllabus had not changed, but “\emph{when it comes to the placement prep […] sometimes the free tools are not sufficient.}” Paid models gave “\emph{way better quality output than a free tier model,}” according to P10, yet few participants paid. P4 explained, “\emph{[The paid version] helps you way more than the free tiers. But as students, we may not have the budget for that,}” and P12 put subscriptions to the newest models at 15,000 to 20,000 rupees, well beyond what students could pay. The gap therefore fell where participants placed the most value.

Language added a further layer for some students. According to P13, a classmate who was “\emph{not good with English}” used chatbots in Tamil but “\emph{has to, like, put in more effort to explain it to the chatbot,}” while P22 practised spoken English with Gemini daily, something classmates did not do. These accounts come from few participants, but they suggest that AI served best those who could already use it in English.

In preparing for placement, participants did much of the work themselves. They worked out what the gate rewarded, built tools to rehearse it and paid for better models where they could. Faculty, as Section~\ref{sec:t4} shows, more often decided whether AI could be used than taught how to use it.

\subsection{Clearing, Not Learning}\label{sec:t3}

The coursework that fed the CGPA was where participants used AI most routinely, and where they were most candid that they used it to \emph{clear}, that is, to pass assessments, collect their marks or submit work on time. They seldom presented this as a loss of learning. Instead, they saw it as a reasonable response to an assessment system that, in their account, had emphasised recall long before generative AI. They decided task by task whether understanding was worth the effort, weighing whether it would be checked, whether AI could be trusted and whether the effort would pay off later, often in placement. AI also served as a tutor where teaching fell short, and much of the learning participants valued happened with its help.

\paragraph{The exam rewards a reproducible answer, and AI produces one}
Participants said end-of-semester examinations were closed-book, handwritten and marked against the professor’s notes. As P13, a data science student, explained,

\begin{quote}
\emph{They are expecting the answers to be what they tell. […] If you’re using the same keywords, the same subtopics as mentioned in the notes, you are good to go.} (P13)
\end{quote}

Because answers of this kind could be prepared in advance, examination preparation became the main use of AI for study. Preparing in the last day or two before an examination was already common, and AI made the practice faster and more routine by summarising material and predicting what was likely to be asked (P1, P2, P4–P7, P12, P13, P18, P19). P17 uploaded their notes and asked ChatGPT for the important questions and answers, reporting that “\emph{most of the time, uh, the questions given by chat GPT will appear in my examination,}” and P22 requested answers already sized to the mark scheme, “\emph{If we want only two-mark, two-mark; if we want only five-mark, five-mark.}”† In an overnight session before one examination, P13 uploaded the syllabus, their notes and their professor’s PDF and told the model to strip out the explanations:

\begin{quote}
\emph{I do not want the extra parts where I understand this stuff. I want the topics to be explained in such a way that I can just easily, like, mug up and write it. […] I made sure to cut down the explanation part and just the, like, exam-ready answers part.} (P13)
\end{quote}

P2 used AI to find questions repeated across fifteen years of past papers and then saw “\emph{those exact questions being repeated}” in the examination. P2 also mentioned peers who used AI only to reach the pass mark, asking it to “\emph{go through the past 15 years, tell me what’s likely to get me like 40 on 100, because that’s my pass mark. […] I think it’s really killing the curiosity.}” Memorisation without understanding predated AI, as P14 acknowledged, but in participants’ telling AI lowered its cost, since the expected answer could now be generated, trimmed to the marks and rehearsed overnight.

\paragraph{Assignments triaged by their perceived value}
Assignments were triaged in a similar way, by their perceived value (P2, P3, P5, P6, P12, P13, P18). Some, P12 recalled, were set “\emph{just for the sake. Just write this,}” and from those “\emph{you don’t learn anything from it. So, you just want to copy-paste the whole thing from it.}” Some faculty shared this view. P5 recalled a professor telling the class, “\emph{it’s 20 marks. You have AI at your hand, just use it and get those 20 marks and get your internals full.}” Deadlines pushed these judgements further. P2 called AI “\emph{a great get out of jail free card for a lot of us}” for moments when “\emph{it’s just physically not possible for you to probably study the concept and then answer whatever is given to you,}” while acknowledging that relying on it this way “\emph{completely takes away from the learning factor}”. P13 put the trade-off directly: “\emph{using AI for assignment is good for a deadline, but not good for you.}”

\paragraph{Effort reserved for where AI fails or understanding is checked}
Participants did the work themselves in two kinds of situation, and they framed both as protecting their marks as much as their learning. Several kept mathematics and numerical work away from AI because its answers were often wrong (P4, P10, P12, P15, P17, P20, P23). Numerical examinations award marks for each step of working, and P4 argued that such subjects therefore require genuine understanding, since copying an answer is something “\emph{you can’t do […] in an exam.}” Participants also invested in their own understanding when they expected to explain their work in person (P1, P6, P8, P9, P13, P21, P23). P13 avoided AI for presentations because “\emph{when I’m getting asked a question, I would never be able to answer it.}” Some professors made this check explicit. P1 recalled professors who permitted AI on assignments, asked which prompts had been used and then held a viva, and P23’s professors told students, “\emph{Use whatever you want, but you should know it. You shouldn’t just blindly mug up and come.}”†

\paragraph{Where AI helped them learn}
AI did not only help participants clear. In a system that, in their accounts, did little to support understanding, it was also how many of them learned. The clearest reason they gave for turning to AI was that teaching assumed knowledge they did not have. P21 spoke of professors who began each class with “\emph{a mindset}” that students “\emph{will already know this; they will have already studied this,}” although “\emph{not all students know even the basics of the topic being taught. That is why all students suffer.}”† AI, by contrast, “\emph{doesn’t teach you in a way that you’re already a PhD scholar. It teaches you from a basic child,}” as P12 put it. Having come to computer science from a biology and mathematics background, P11 felt like “\emph{the only one in the classroom who does not understand it very clearly,}” and AI “\emph{teaches me from the beginning on what to use well and what all syntax I’m supposed to use.}” AI was also always available. P14 called it “\emph{my personal professor,}” since the actual professor could not be called “\emph{at midnight,}” whereas AI could be asked “\emph{for 100th time}” and “\emph{will definitely answer.}”

Some participants turned this tutor toward understanding as well as marks. Current students used AI to test themselves, solving past papers and checking their answers against AI (P16), having it set a mock question paper the night before an examination (P19) or generating progressively larger practice problems at a professor’s suggestion (P20). P7 grounded AI in sources they trusted, since “\emph{when you give the resource to AI, for example, a book reference or a research paper reference, and then ask it to break it down, explain particular concepts, then you get more comfort and more confidence.}” P21 summarised the dual value of these practices: “\emph{Both understanding and marks, ma’am. Because AI is quite useful for both.}”†

Much of the learning participants valued most happened beyond the assessed curriculum, and AI was central to it. P18 found AI more useful than a Google search when learning “\emph{something we know nothing at all about,}”† and relied on it to learn the coding and data structures that placement required (Section~\ref{sec:t2}). P14 called AI a “\emph{free guide}” that helped them pass a placement coding test. Graduates continued to learn this way at work, asking models to explain unfamiliar topics as if to “\emph{a beginner}” (P5) or using AI daily “\emph{Just to serve my curiosity}” (P15). Learning was not always the outcome, however. On placement problems, P18 admitted, they would “\emph{just copy paste}” the AI’s solution “\emph{without thinking on my own how that solution can be done.}”†

\paragraph{Two tracks, scoring and learning}
Participants’ strategic use of AI was a response to a system they saw as outdated and narrowly measured. In P7’s words, “\emph{the same books are followed from like 50, 60 years. The system is not changed,}” and P6 was examined on “\emph{Some 1980s stuff}” before being asked about current technologies in job interviews. At the same time, marks determined access to opportunity, and for P5, “\emph{GPA matters a lot for everything and marks matter a lot for everything.}” There were times, P9 said, when “\emph{understanding of subject matter goes secondary and, uh, actually being able to do, uh, the job that is required, that is give the answer in the way that the professor expects, uh, is probably the most important thing.}”

This pattern was not universal. P7’s institute set open-book tests in which any resource was allowed, yet “\emph{you can’t answer most of the questions,}” P9’s provided formula books so that students “\emph{could just apply things right in the exam itself,}” and questions that professors wrote themselves resisted AI (P20). These exceptions were, however, confined to particular courses or professors, and they often coexisted with the dominant pattern within the same institution. P9, whose institute provided the formula books, identified laboratory sessions, in which code was copied by hand into lab records, as the place where AI-generated work was most common.

Faced with marks that mattered and a curriculum that seemed beside the point, participants ran scoring and learning as separate tracks. AI carried much of the first and supported much of the second, which took place largely outside the assessed curriculum, in preparation for placement and work. When the interviewer suggested that AI let students find a loophole instead of cheating, P5 rejected the softer term: “\emph{No, I would say cheat the system. Definitely it cheats the system because the system is only based off of one thing, which is marks and that’s it.}” Responding to the interviewer’s summary, P13 put it most simply: “\emph{So, the college expects a structure. The AI also gives us a structure. That’s all. So, we clear the exams but not the learning part. We never get it.}”

\subsection{Answering Upward, Sharing Sideways}\label{sec:t4}

Two relationships shaped how participants used AI, and they ran in different directions. Upward, toward faculty, the relationship was steep. Professors set the questions, awarded the marks and decided, course by course and often without saying so, whether AI could be used. Many accepted it, but few changed how they taught or assessed in response, and where institutions acted at all, it was through detection. Sideways, among peers, the relationship was flat and open, and much of what participants knew about AI came from one another.

\paragraph{Deference follows the grader}
Most participants trusted their professors over AI. Asked whom they would believe when AI and a professor gave different answers, 15 of 23 chose the professor or the textbook (P1, P2, P4, P6, P8–P10, P13–P16, P19–P21, P23), usually because of the professor’s experience. For some, however, trust followed the person who graded, not the person most likely to be right (P1, P8, P9, P13). As P8 put it, “\emph{I believe the professor more. Because, again, the professor is the one giving me the marks, right?}” P13 likewise concluded that “\emph{trusting a professor in an exam point of view is much better because he’s the one who will be correcting our papers.}” For P9, a computer science graduate, “\emph{especially in the Indian education system, I feel like whatever the professor says goes,}” so when AI solved a problem differently from the way professors expected, P9 prompted it to follow the expected format, even when this was less efficient.

\paragraph{Little room to argue, and rules left unstated}
Participants saw few ways to contest this authority. P6 recounted a viva in which a teammate mentioned ChatGPT, after which the professor called in other faculty, mocked the group and demanded an apology letter from the students’ parents. P6 felt unable to object. “\emph{I can’t say much because, end of the day, if I say something and [he] takes it to his heart, I think I’m screwed for the rest of the semester.}” According to P18, some professors asked students to explain code and, if they could not, accused them of using AI and scolded them.

In the absence of institutional rules, each professor set their own, and participants saw faculty as divided (P1, P2, P4, P7, P17). Some encouraged AI while others treated it as an enemy (P7), and between these ends lay a large space in which professors said nothing (P2, P9, P15, P16, P19). Participants read this silence in opposite ways. P9 took it as tacit acceptance, whereas P2 and P16 read it as a risk and concealed their use. P2 explained:

\begin{quote}
\emph{So even if they’re not explicitly mentioning that AI use is not allowed, if they were to ever ask me or ask any of my classmates, our instant response is going to be like, you know, no, we did not use AI because of how reluctant they are and how that’s probably gonna, you know, maybe change the way they’re grading our reports and assignments.} (P2)
\end{quote}

\paragraph{Acknowledged, not incorporated}
Many professors did accept AI, and participants reported acceptance as often as bans. Some encouraged it outright. P5 recalled professors who “\emph{explicitly said you have it, please use it, get full marks,}” and P22’s professors told students to “\emph{put it in ChatGPT, guys.}”† Others accepted it with a condition. P21’s professors told students that “\emph{your using AI is not a problem. But have you understood the task you do using it?}”† Over participants’ degrees, positions shifted mostly toward acceptance (Section~\ref{sec:t1}). Acceptance, however, rarely went further. Even professors who allowed AI seldom showed students how to use it or changed how they taught or assessed, and their permission usually came as a caution, not as instruction (P1, P2, P4, P14, P17, P21, P23). Across four years, P6 could name only “\emph{one teacher who was essentially saying, yeah, use AI, but make sure you’re learning.}” When the interviewer suggested that professors accepted AI without changing how they taught or assessed, P5 agreed: “\emph{they’re okay with AI, but they’re not actually embracing it.}” A few went further, as when one of P2’s professors had AI explain a concept in class and then showed where its account was wrong, but such cases were exceptions.

\paragraph{Policing without policy}
None of the institutions in participants’ accounts had told students how AI should be used. As P9 put it, “\emph{regarding academic integrity, I guess AI has been clubbed with every other form of cheating.}” Where institutions acted, it was through numerical limits on AI content in submissions (P2, P12, P15), checking software (P6, P9) and network blocks during tests (P4, P11, P18). These measures covered only supervised tests. Tests held online without them were open to AI, and two participants said so plainly: “\emph{Some would also use AI to cheat in online examinations}” (P4), and “\emph{In terms of online assessments, you can cheat using AI}” (P5). P2’s university set a limit of about 5\% on its AI checker for reports, and P12 read a limit as an allowance more than a prohibition: “\emph{they want us to have, like, 30\% AI used.}” Participants did not trust the checkers, which, as P15 had seen, erred in both directions, and they met detection with a repertoire of evasion (P1, P3, P4, P8, P13, P15, P17, P18). The most common technique was to “\emph{humanize}” AI output, which P3 did routinely “\emph{so that it passes through the AI detector.}” P8 asked the model to “\emph{humanize it, make it sound more Indian,}” which suggests an understanding that detection depended on how text sounded, and that sounding like an Indian student was part of sounding human.

The only checks participants saw as reaching understanding were oral (Section~\ref{sec:t3}), and even these could be anticipated. When AI wrote P8’s code, “\emph{I try understanding it so I can explain it to the teacher if they do ask me,}” P8 explained, before summing up, “\emph{I try to hide it, basically.}” Asked what they would change, few participants called for stricter detection. Most proposed changing what was assessed, toward application-oriented questions (P12) and, in P7’s agreement with an interviewer’s suggestion, assessments that permit AI but cannot be answered by it alone. P9 proposed testing the two separately, returning to “\emph{pen and paper exams to ensure students actually know what they’re doing}” while using online assessments to test how well students applied AI, because “\emph{these things have to be tested in different kinds of ways.}”

\paragraph{Peers teach one another how}
Among peers, AI use was open and routine. Many participants shared prompts, outputs or entire conversations with classmates (P1, P5, P7, P8, P13, P18, P20–P22), and sharing peaked before examinations. In P8’s class, the class representative consolidated everything students needed for an examination into a single ChatGPT chat an hour or two before it began and shared it with everyone, and P18’s group posted their questions into one shared conversation so that everyone read the same answers. Prompts circulated as a way to reproduce good results. As P21 explained, “\emph{if the prompt I entered was better than theirs, I share my prompt with them, and we use that prompt to derive the same output.}”† Knowledge of tools travelled the same way, as classmates passed on news of model releases and a working sense of which model suited which task (P7, P11, P19).

Because adoption spread informally, it spread unevenly (P4, P7). Given the same task, P7 observed, some classmates “\emph{have already done the task with AI and some people don’t know that this can be done,}” and P4 called using AI “\emph{also a skill}” that some classmates had not developed. Views on whether faculty could fill this gap were mixed. Eight participants wanted guidance from professors or their university (P1, P3, P7, P11, P13, P21–P23), while five saw little value in it (P5, P6, P8, P10, P12), among them P5, who said, “\emph{I don’t think we rely on the professors for anything in the first place.}” The split appeared to follow participants’ confidence with AI. Those who rejected guidance saw themselves as capable users, among them P10, who said, “\emph{we are the next generation and we’re pretty smart and, you know, we’re growing up with AI,}” whereas those who wanted it more often spoke of dependence or inexperience, as P11 did in feeling that guidance would have prevented becoming “\emph{that addicted to like asking AI.}” Professors decided whether AI could be used, often without saying so, while peers taught one another how. No one, in participants’ accounts, taught them what to learn for work.

\subsection{Ready for What?}\label{sec:t5}

Participants were candid about what this arrangement left them ready for. AI had become their first tutor, and for many it had also become the first response to any task. Many spoke of a dependence they had not chosen, of thinking skills they felt they were losing and of AI-assisted work they could not fully claim as their own, and several connected these worries directly to the jobs they sought.

\paragraph{Dependence before thinking}
The tutor that helped participants learn (Section~\ref{sec:t3}) could also replace their thinking. The participants for whom teaching had left the widest gaps, such as P11 and P21, were among those who said they depended on AI most, and for many participants AI had become the first response to any task, however small (P1, P3, P4, P5, P7, P14). P7 recognised the same reflex in themselves, typing every task into the chat window without thinking it through, with “\emph{almost loss of struggle,}” and P14 applied it even to problems they knew they could solve, saying, “\emph{It’s simple, but I’m not ready to think.}” Some used AI to confirm answers they had already reached (P2, P11, P12, P17). “\emph{Even if I know the answer, to be frank, I will not trust myself,}” P17 said, and P2 admitted to “\emph{seeking reassurance}” when feeding a plan written on paper to AI.

Many participants felt they were losing their own thinking (P1, P2, P5, P6, P9, P14, P20, P21). P2 gave the fullest account:

\begin{quote}
\emph{I think it takes away from our ability to actually sit down, think with it. You know, a lot of the times I’m not even convinced with things I’m studying. And that’s where it’s like, you know, like, convince me, like, I want to be convinced. And now with AI, it’s just like, it’s literally like shooting an arrow straight into both. […] It’s a direct, this is what you asked for. This is what you take. […] In doing that, I feel like I’m losing my ability to just sit down and think.} (P2)
\end{quote}

P2 concluded, “\emph{I think in using AI, I’m shrinking my capacity to learn,}” and P21 saw themselves as now “\emph{at a stage where I understand only by depending on AI.}”† P5 felt that their ability to approach problems “\emph{in terms of like a coding perspective has deteriorated,}” and P9, now a junior developer, observed that seniors who had studied before AI were “\emph{much better}” at problem solving and decision making than their own batch. Participants also named specific skills they had lost, from basic tasks such as writing a SQL query or an email (P7) to writing, typing and calculation (P11, P13, P15, P18, P23). Not all participants felt diminished. P10 argued that delegating routine work to AI raised the level at which students worked without removing skills, and P9 felt that AI-assisted debugging had made them more knowledgeable, not less.

\paragraph{Work they could not fully claim}
The costs participants tied most directly to work concerned ownership. Completing work with AI let P7 “\emph{sleep peacefully that the work is completed,}” but when it came to discussing it, “\emph{I do feel that I may not be able to answer all the questions thrown to that assignment or the report or the presentation.}” P7 had seen the same among juniors and friends. A student who had used AI on a project felt “\emph{some level of underconfidence}” when presenting it in a placement interview, thinking, “\emph{This is not wholly done by me.}” In P7’s words, “\emph{you don’t go with the same ownership confidence that you used to do.}” P14 feared for their livelihood more directly: “\emph{If there is no [AI], then what can I do? To be honest, if there is no [AI], there is no job for me.}”

Students who had entered college with AI already in place (P11, P13, P16, P21, P23) framed these concerns in terms of what lay ahead. P21, in their third year, warned unprompted that students who depended on AI “\emph{from the beginning of the year, the college year}”† would struggle with employers, because, having let AI do the work, “\emph{I can’t meet the company requirements.}”† Their advice was to use AI less, not to avoid it.

\paragraph{Literacy learned through use}
What participants knew about using AI well, they learned through trial and error, from friends and from internships, not from their institutions (P3, P5, P9, P16, P18, P20, P23). They developed rules of thumb for when to trust it. P6 drew the line at stakes, saying, “\emph{If there’s a stake on it, make sure you are, like, using your skill, your brain, and not AI,}” and P20 had learned to distrust the model’s readiness to agree, since a challenged answer drew the reply “\emph{Correct. I’m the one who made the mistake.}”† This literacy was unevenly distributed. P2 acknowledged that classmates with stronger coding skills were “\emph{using AI more efficiently than I am,}” while P5 attributed their own effectiveness to experience: “\emph{I know where it goes wrong, where it starts hallucinating.}”

The literacy participants built helped them prompt and catch some of AI’s errors, but it did not extend to managing their own reliance. Readiness for AI-first work was something they assembled for themselves, unevenly and often with misgivings, from the gaps their institutions left.

\section{Discussion}\label{sec:discussion}

We first return to our research questions. For RQ1, participants saw themselves heading into workplaces that expected AI-first work, and they prepared for that transition mainly through their own, AI-coached placement preparation and learning they pursued outside the assessed curriculum, while worrying that dependence and weak ownership of AI-assisted work would leave them unready. For RQ2, they used AI to clear coursework and examinations that rewarded reproducible answers, to learn where teaching fell short and to prepare for placement screens that mostly excluded it. For RQ3, institutions shaped this use mostly through what they assessed, not through policy, which no participant could point to, and also through professors whose positions on AI often went unstated, through detection and through access left to company promotions.

Across the five themes, a single pattern recurs. The gates participants passed through, from semester examinations to placement screens, were built before generative AI and still largely assumed its absence, while the work beyond them assumed its presence. We discuss three consequences of this mismatch. We argue 1) that AI-assisted clearing weakens what grades and screen scores signal, 2) that AI-free gates leave both students and employers without a shared account of what readiness for AI-first work requires, and 3) that in the space between them, readiness becomes a burden students carry on their own. We then turn to implications for employers, institutions and tool designers.

\subsection{When Clearing Produces the Signal}

Recruiters screening large cohorts in a placement season rely on signals such as a CGPA above a cut-off or an aptitude test score, which are assumed to be harder to obtain for weaker candidates ~\cite{10.2307/1882010} (Section~\ref{sec:signal}). Our participants’ accounts suggest that AI erodes this assumption at several points along the path. In coursework, examinations rewarded answers that could be predicted and rehearsed, and AI made the prediction and rehearsal cheap. Participants used AI to forecast likely questions, to compress chapters into answers sized to the mark scheme and to keep their CGPA above the cut-offs that companies set, often while describing the result, in P13’s words, as clearing the exams “\emph{but not the learning part.}” Online tests that formally excluded AI, both in college and at the placement stage, were open to it in practice. Participants reported that students used AI to cheat in them (P4, P5), and by one estimate this was routine in placement tests (P8). If a CGPA can be maintained, and a screen passed, with substantial help from AI, then the signal becomes cheaper for everyone to acquire and tells employers correspondingly less about the skill it was meant to indicate.

This reading extends Corbin et al.’s argument about assessment validity~\cite{10.1080/02602938.2025.2503964}. Closed-book, invigilated, handwritten examinations are structural safeguards that AI cannot enter, and Joshi et al.’s instructors expected such examinations to keep their significance~\cite{joshi2023withgreatpowercomes}. Yet because the examinations rewarded rehearsable answers, participants used AI upstream, during preparation, so that the safeguard secured the conditions of assessment without securing its object. The same structure reappears at the placement gate, in invigilated or proctored tests whose content can be prepared for, or circumvented, with AI. A signal protected only at the moment of measurement can still be produced elsewhere.

Employers’ responses, as participants experienced them, may compound the problem. When a signal weakens, the theory predicts that employers will demand stronger ones, and participants reported interviews that had become “\emph{a lot harder than what the actual job might be}” (P9), dominated by DSA problems they did not expect to use. Harder screens of the same kind raise the cost of the signal without restoring its link to the job, and they intensify the AI-coached preparation that produced the problem. Armstrong and Metaxa found that, among U.S. computer science students, referrals and household income predicted job offers while more egalitarian strategies did not~\cite{10.1145/3711038}. Our data cannot test this in India, but participants’ observations that free tiers fell short for placement preparation (P7) and that few students could pay for better models (P4, P12) suggest one way in which money could re-enter selection as test-based signals lose force. We offer this as a hypothesis for future work.

The signal also matters to students themselves. Several participants could no longer tell how much of their own work they understood, and some checked answers they had already reached for reassurance. A grade earned with AI’s help tells its holder as little as it tells an employer, and the loss of ownership P7 observed among students presenting AI-assisted projects in placement interviews suggests that some students feel this gap most sharply at the moment they are asked to account for their work.

\subsection{AI-Free Gates for AI-First Work}

Participants were caught between two systems. Coursework and placement screens largely assessed unaided skills, while the workplaces graduates described expected AI-first work on compressed deadlines. Participants delegated to AI tasks that once built foundational skill, such as writing a basic query or an email, and for some graduates learning had become secondary to delivering results. Yet they had received little guidance on what to learn in place of what they delegated. P10 argued that AI raised “\emph{that level}” of what students should be capable of, and P5 that base knowledge remained necessary to use AI well, but in participants’ accounts neither their institutions nor, in most cases, their professors had defined what that level was. Employers’ screens did not define it either, since they mostly tested candidates as though AI did not exist.

This gap mirrors what Yu and Moon call \emph{Absorption}, in which GenAI takes over entry-level work and removes the productive struggle through which juniors once built expertise~\cite{yu2026seniorgenerativeaierodes}. Their account centres on the workplace. Ours suggests that the same loss begins earlier, in a degree whose assessments could be cleared with AI and whose placement screens rewarded preparation more than practice. Lee et al. warn that skills left unpractised on routine tasks may deteriorate~\cite{10.1145/3706598.3713778}. Our participants’ declining confidence in their own reasoning, and the work they could not fully claim as their own, suggest what this may look like for students who have not yet built those skills, and who are therefore poorly placed to evaluate what AI produces for them.

Participants’ accounts also point to a skill that neither gate assessed, the ability to account for AI-assisted work. The checks they experienced as reaching understanding were oral, and the costs they tied most directly to work concerned explaining, defending and owning work produced with AI. In workplaces where much code and many documents begin with AI, the capacity to evaluate and stand behind that output may be closer to the core of the job than the unaided recall that most screens still measure.

These questions have particular weight in India. India is reported to be ChatGPT’s second-largest user base and to have more student users than any other country~\cite{techcrunch2026india}, and its technology sector accounts for a large share of the world’s outsourced technology services~\cite{nasscom2025strategic}. P7 sketched the division of labour this implies, in which multinational firms keep architecture and problem framing and pass routine work to Indian IT companies, and routine work is precisely what AI absorbs first. Participants took for granted that AI belongs in their education. The open question is where graduates will contribute alongside it, and answering that question means both reforming what institutions and employers assess and bringing AI into it. Doing only the second would make the existing gates easier to pass, and doing only the first would leave students preparing for work that no longer exists in the form the gates assume.

This question became pressing within a single degree (Section~\ref{sec:t1}), and it is unlikely to resolve itself as AI becomes familiar, since students who entered college with AI already in place raised some of the clearest concerns about whether they could meet “\emph{the company requirements}”.

\subsection{Readiness as an Individual Burden}

In the space between these gates, readiness became something students assembled for themselves. Participants’ placement preparation was often sophisticated, including graded mock interviews, interviewer profiling, custom GPT trainers and overnight consolidation after a shortlist, and their rules of thumb for when to distrust AI were hard-won. Cao et al. argue for recognising what learners already bring to AI literacy education~\cite{10.1145/3706598.3713173}, and these strategies are a resource of that kind. But they were unevenly distributed, and the conditions that produced them placed a heavy load on the individual.

Faculty authority added to that load. Adnin et al. documented U.S. students adjusting how carefully they hid AI use to each instructor’s perceived strictness~\cite{10.1145/3706598.3713393}. Our participants made the same adjustments within a far steeper hierarchy, in which professors set the rules course by course, often without stating them, and in which the cost of misjudging a professor fell on marks that later decided placement eligibility. In this setting, silence did not read as permission, and calibration settled into concealment by default. At the same time, knowledge about AI travelled sideways. Where Hou et al. found that AI-mediated help-seeking left North American students isolated and ashamed in front of peers~\cite{10.1145/3724363.3729024}, and Ling et al. traced under-reporting to embarrassment~\cite{10.1145/3772318.3791073}, our participants shared prompts, chats and model knowledge openly, and learned from one another what their institutions did not teach.

Peer networks, however, spread AI use unevenly, and so did money, language and confidence. The participants who appeared to navigate most effectively saw themselves as motivated to learn beyond the syllabus and confident enough with AI to know when to stop, and some could pay for better tools. They rarely wanted guidance and had few complaints about its absence. Those who wanted it included all four participants from the arts-and-science college, a first-year student who had come to computer science from biology and a civil engineering graduate from a smaller college, and they more often spoke of dependence, of tools they could not afford or of struggling to say what they wanted from a model (Sections~\ref{sec:t2} and~\ref{sec:t4}). As P23 said, “\emph{what I’m expecting, only I know. I can’t express that to it.}”† Language may widen these gaps. Classmates who used chatbots in Tamil had to work harder to be understood (Section~\ref{sec:t2}), which fits Verma et al.’s observation that tools built on models trained mainly in English serve other languages less well~\cite{10.1145/3736645} and Agarwal et al.’s concern about quality-of-service harms, in which non-Western users must work harder for similar benefits from AI~\cite{10.1145/3706598.3713564}.

These patterns are also consistent with Ma et al.’s finding that prior experience continued to predict success with LLM assistance~\cite{10.1145/3772318.3791283}. Our data cannot separate these factors, and we offer them as hypotheses. What they suggest is that, when institutions and employers leave readiness undefined, the students who arrive best equipped are the ones most likely to acquire it. Students who choose to learn beyond the curriculum are likely to do better, which is a reasonable return on personal effort. An arrangement that depends on such effort is another matter, and closing the gap should not fall to the students who feel it most. Most students cannot be expected to arrive knowing how to avoid over-dependence, when to stop using AI or what to learn instead, and these questions should not be left to them alone.

\subsection{Implications}\label{sec:implications}

\textbf{For employers and placement processes.} Our findings suggest that placement screens could be brought closer to the work behind them. First, screens could assess how candidates work with AI as well as what they know without it. Participants welcomed rounds that allowed Copilot or asked candidates to build an application with AI, as “\emph{more of a reflection of what’s going on in real life}” (P8), and doubted that AI-free online tests could be held. Second, because AI makes familiar problems and their answers cheap to rehearse, recalling a practised solution says little about readiness. Problem solving, including DSA, could instead be tested in ways that acknowledge AI, for instance by asking candidates to explain, critique and extend a solution produced with it, which would also reward the ownership that participants felt they were losing. Third, employers could treat CGPA cut-offs and online test scores as weaker evidence than they once were, and weigh them alongside work samples and conversations. Finally, employers are well placed to tell institutions what readiness for AI-first work involves, and our participants’ accounts suggest that no one in the current arrangement is telling students. Clearer statements of expected skills, in job descriptions and in placement briefings, would give students and faculty a target to prepare for.

\textbf{For institutions and educators.} The most useful changes are likely to be local and structural. First, professors could state their position on AI for each course and each task, since participants read silence in opposite ways and it pushed the more cautious toward concealment. Second, assessment could reward reasoning over reproduction, through vivas, explanations of submitted work, application-based questions and open-book tasks that permit AI but cannot be answered by it, alongside examinations that test unaided understanding. Third, guidance on AI is most valuable when it is specific to a subject and connected to the work students are preparing for, and when it begins from the basics that teaching often assumes. Fourth, faculty need time and training of their own if they are to model the use they expect. Finally, institutions could negotiate AI access for their students, instead of leaving the quality of placement preparation to the offers students can claim or the subscriptions they can afford. These changes are easier to propose than to make. They require AI access for students across very different institutions, AI literacy for faculty as well as students, and changes to curricula and assessment that are slow to make, and participants, who see mainly the consequences, may underestimate these constraints. The changes are needed nonetheless, since in the meantime it is students who bear the cost.

\textbf{For AI tool design.} Tools could ask students to commit to their own answer before seeing the model’s, a form of the cognitive forcing that Raees et al.’s review reports can reduce over-reliance~\cite{10.1145/3772318.3791467}, and could make the provenance of AI-assisted work visible to the student, which may support the ownership participants felt they lacked. Sycophantic agreement, which most novices in Bo et al.’s study did not notice~\cite{10.1145/3772318.3791365}, is particularly costly for students preparing for screens and jobs in which they will have to defend their answers.

\section{Limitations and Future Work}\label{sec:limitations}

Our sample (Section~\ref{sec:participants}) limits how far these findings travel. Students at less-resourced institutions, at rural colleges and in regional-language-medium programmes were not represented, and the inequalities we observed may be sharper there. With only two women in the sample, our findings largely reflect men’s experiences. Graduates’ accounts of college were retrospective and may have been shaped by their later work, and all accounts were self-reported, so participants may have under- or over-stated their own AI use. Some nuance may have been lost in translating the Tamil–English interviews. Finally, we did not interview faculty, administrators, recruiters or managers, so their positions appear here only through students’ perceptions, at a time when tools, offers and policies were changing rapidly.

Several directions follow. Interviews with recruiters and hiring managers would show how employers interpret CGPA and screen scores now that candidates prepare with AI, and whether the weakening of signals we infer from students’ accounts is visible from the other side of the gate. Longitudinal work following the cohort that entered college with AI already present would show how their concerns about readiness develop as they reach placements and work. Larger-scale surveys could test whether the patterns identified here hold across institution tiers, regions and disciplines, and intervention studies could evaluate the assessment and placement designs participants proposed, such as AI-permitted build tasks paired with oral explanation.

\section{Conclusion}\label{sec:conclusion}

We interviewed 23 engineering and computing students and recent graduates in India about the place of AI on their path from coursework to work. In their accounts, the gates along that path, from semester examinations to placement screens, were built for a world without AI, while the jobs beyond them expected AI from the first day. Participants used AI to clear the gates, to prepare for them and, where teaching fell short, to learn, often with considerable skill and mostly on their own initiative, under professors whose stance on AI was seldom explicit and with tools whose quality depended on what they could afford. Drawing on signaling theory, we have argued that AI-assisted clearing weakens what grades and screen scores certify, for employers and for students themselves. As AI-first work arrives in India’s technology sector, employers and institutions face a shared task, which is to define what readiness for that work requires and to assess it directly. Our participants are already attempting this on their own.

\section*{Disclosure of AI Use}
The authors conducted all interviews and the literature review, and made all analytic decisions, including the coding, the themes and the interpretations presented in this paper. We used Claude Opus 5.5 (Anthropic), a large language model, as a writing and organisation aid. It helped collate interview excerpts, draft prose from the authors’ notes and outlines, edit text for grammar, clarity and structure, check quotations against the transcripts and format the paper. The authors reviewed, revised and verified all AI-assisted material and take full responsibility for the content of this paper.

\begin{acks}
We thank our participants for their time and candour. This work was conducted as part of the Lumiere Individual Research Program.
\end{acks}

\bibliographystyle{ACM-Reference-Format}
\bibliography{main}

\appendix

\section{Participants}\label{app:participants}

Institutions are described by type, and gender is omitted to protect anonymity. Programmes and years are as participants described them (Table~\ref{tab:participants}).

\begin{table}[tp]
  \caption{Participant attributes.}
  \label{tab:participants}
  \small
  \begin{tabularx}{\textwidth}{@{}ll>{\raggedright\arraybackslash}Xll@{}}
    \toprule
    \textbf{ID} & \textbf{Institution type} & \textbf{Programme} & \textbf{Status} & \textbf{Interview language} \\
    \midrule
    P1 & Private university & B.Tech., Computer Science and Engineering & Current, 3rd year & English \\
    P2 & Private university & B.Tech., Biotechnology & Graduate & English \\
    P3 & Private engineering college & B.E., Civil Engineering & Graduate & English \\
    P4 & Private university & B.Tech., Mechanical Engineering & Graduate & English \\
    P5 & Private university & B.Tech., Biomedical Engineering & Graduate & English \\
    P6 & Private university & B.Tech., Electronics and Communication Engineering & Graduate & English \\
    P7 & Centrally funded technical institute & Dual degree, Electrical Engineering and M.Tech. in Machine Learning & Graduate & English \\
    P8 & Private university & B.Tech., Computer and Communication Engineering & Graduate & English \\
    P9 & Private university & B.Tech., Computer Science & Graduate & English \\
    P10 & Private university & B.Tech., Computer Science & Current, final year & English \\
    P11 & Private university & B.Tech., Computer Science and Engineering (AI and ML) & Current, 1st year & English \\
    P12 & Private university & B.Tech., Mechatronics & Graduate & English \\
    P13 & Arts-and-science college & B.Sc., Data Science and AI & Current, 3rd year & English \\
    P14 & Public state university & Integrated M.Sc., Information Technology & Current, 5th year & English \\
    P15 & Centrally funded technical institute & Dual degree, Mechanical and Computational Engineering & Graduate & English \\
    P16 & Public state university & Integrated M.Sc., Information Technology & Current, 4th year & English \\
    P17 & Public state university & Integrated M.Sc., Computer Science / Information Technology & Current, 5th year & English \\
    P18 & Public state university & Integrated M.Sc., Computer Science / Information Technology & Current, 5th year & Tamil–English \\
    P19 & Public state university & Integrated M.Sc., Computer Science / Information Technology & Current, 4th year & Tamil–English \\
    P20 & Public state university & Integrated M.Sc., Computer Science & Current, 4th year & Tamil–English \\
    P21 & Arts-and-science college & B.Sc., Data Science and AI & Current, 3rd year & Tamil–English \\
    P22 & Arts-and-science college & B.Sc., Data Science and AI & Current, final year & Tamil–English \\
    P23 & Arts-and-science college & B.Sc., Data Science and AI & Current, 3rd year & Tamil–English \\
    \bottomrule
  \end{tabularx}
\end{table}

\end{document}